\documentclass{llncs}

\IfFileExists{macpath.tex}{\input{macpath.tex}}{} 
\usepackage[hyperref=auto,backend=bibtex]{biblatex}
\bibliography{kwarc}

\newcommand\tabhead[1]{\small\textbf{#1}}
\renewcommand\baselinestretch{.996}

\usepackage{graphicx}
\usepackage{stex-logo}
\usepackage{paralist}
\usepackage{comment}
\usepackage{wrapfig}
\usepackage{amstext}
\usepackage{xcolor}

\usepackage[hide]{ed}
\usepackage[final]{svninfo}
\svnInfo $Id: paper.tex 362 2012-04-20 06:19:01Z kohlhase $
\svnKeyword $HeadURL: https://svn.kwarc.info/repos/sissi/doc/papers/mkm12/paper.tex $

\def\myCitation#1{``{\emph{\small{#1}}}''}

\def\name#1{{\textsc{#1}}\index{#1@{\sc #1}}}
\def\cellrange[#1,#2]{\textsf{[#1:\kern-.10em#2]}}
\def\cell[#1]{\textsf{[#1]}}
\def\cellval#1{\textsf{[\kern-.15em[#1]\kern-.15em]}}
\def\cellform#1{\textsf{[\kern-.15em[\kern-.15em[#1]\kern-.15em]\kern-.15em]}}
\def\column[#1]{#1}
\def\row[#1]{#1}

\def\symApp{$\mathcal{A}$\xspace}

\def\alex{\texttt{Alex}\xspace}
\def\theo{\texttt{Theo}\xspace}
\def\genericAlex{\mbox{\alex$_{\kern-.3em\text{\symApp}}$}\xspace}
\def\genericTheo{\mbox{\theo$_{\kern-.3em\text{n}}$}\xspace}
\def\ooalex{\mbox{\alex$_{\kern-.3em\text{Calc}}$}\xspace}
\def\msalex{\mbox{\alex$_{\kern-.3em\text{Excel}}$}\xspace}

\usepackage[pdftex]{hyperref}

\def\mathUI{{mSI}\xspace}
\def\rgi{\name{RGI}\xspace}
\def\Math{{\small{{\name{inMATH}}}}\xspace}
\def\noMath{{\small{{\name{noMATH}}}}\xspace}
\def\zbMath{{\small{{\name{infoMATH}}}}\xspace}

\newcounter{patternCounter}
\makeatletter
\newcommand\myPatterns[2][]{\def\@test{#1}%
\refstepcounter{patternCounter}%
\ifx\@test\@empty\else%
\expandafter\xdef\csname pattern@#1\endcsname{\thepatternCounter}\fi%
\begin{center}
  \fbox{\begin{minipage}{.9\textwidth} {\bf{\small{Pattern \arabic{patternCounter}:}}}
      {\myCitation{#2}}
  \end{minipage}}
\end{center}
}
\newcommand\myPattern[1]{\@nameuse{pattern@#1}}
\makeatother

\newcounter{implicationCounter}
\makeatletter
\newcommand\myImplications[2][]{\def\@test{#1}%
\refstepcounter{implicationCounter}%
\ifx\@test\@empty\else%
\expandafter\xdef\csname implication@#1\endcsname{\theimplicationCounter}\fi%
\begin{center}
  \fbox{\begin{minipage}{.9\textwidth} {\bf{\small{Implication \arabic{implicationCounter}:}}}
      {\myCitation{#2}}
  \end{minipage}}
\end{center}
}
\newcommand\myImplication[1]{\@nameuse{implication@#1}}
\makeatother

\newcommand\con[2]{{\small{\texttt{Con\_#1\-\_#2}}}\xspace}

\def\elemZBMathNew{{\small{\texttt{zb}\-\texttt{Math}\-\texttt{New}}}\xspace}
\def\elemZBMathOld{{\small{\texttt{zb}\-\texttt{Math}\-\texttt{Old}}}\xspace}
\def\elemMathSciNet{{\small{\texttt{Math}\-\texttt{Sci}\-\texttt{Net}}}\xspace}
\def\elemArxiv{{\small{\texttt{ar}\-\texttt{Xiv}}}\xspace}
\def\elemGoogle{{\small{\texttt{Goo}\-\texttt{gle}}}\xspace}
\def\elemGoogleScholar{{\small{\texttt{Goo}\-\texttt{gle-}\texttt{Scho}\-\texttt{lar}}}\xspace}
\def\elemCatchup{{\small{\texttt{ar}\-\texttt{Xiv-}\texttt{Catch}\-\texttt{up}}}\xspace}
\def\elemColleagues{{\small{\texttt{my}\-\texttt{Col}\-\texttt{lea}\-\texttt{gues}}}\xspace}
\def\elemOffice{{\small{\texttt{my}\-\texttt{Of}\-\texttt{fice}}}\xspace}
\def\elemFormulaSearch{{\small{\texttt{For}\-\texttt{mu}\-\texttt{la}\-\texttt{Search}}}\xspace}
\def\elemMSCMap{{\small{\texttt{MSC-}\texttt{Map}}}\xspace}
\def\elemVifamath{{\small{\texttt{vi}\-\texttt{fa}\-\texttt{math}}}\xspace}
\def\elemTIB{{\small{\texttt{TIB}}}\xspace}
\def\elemMathoverflow{{\small{\texttt{math}\-\texttt{over}\-\texttt{flow}}}\xspace}
\def\elemResearchGate{{\small{\texttt{Re}\-\texttt{search}\-\texttt{Gate}}}\xspace}
\def\elemBibliography{{\small{\texttt{Bi}\-\texttt{blio}\-\texttt{gra}\-\texttt{phy}}}\xspace}
\def\elemLibrary{\small{{\texttt{my}\-\texttt{Li}\-\texttt{bra}\-\texttt{ry}}}\xspace}

\title{Search Interfaces for Mathematicians}
\author{Andrea Kohlhase}
\institute{Jacobs University Bremen and FIZ Karlsruhe}

\begin{document}

\maketitle

\begin{abstract}
Access to mathematical knowledge has changed dramatically in recent
years, therefore changing mathematical search practices. Our aim with
this study is to scrutinize professional mathematicians' search
behavior. With this understanding we want to be able to reason why
mathematicians use which tool for what search problem in what phase of
the search process.
 To gain these insights we conducted 24 repertory grid interviews with
 mathematically inclined people (ranging from senior professional
 mathematicians to non-mathematicians). From the interview data we
 elicited patterns for the user group ``mathematicians'' that can be
 applied when understanding design issues or creating new designs for
 mathematical search interfaces.
\end{abstract}

\section{Introduction}
Mathematical practices are changing due to the availability of
mathematical knowledge on the Web. This paper deals with the question
whether mathematicians have special needs or preferences when
accessing this knowledge and if yes, what are those? In particular, we
focus on how mathematicians think of search on the Web: what are their
cognitive categories, what kinds of searches do they distinguish, and
which attributes do they associate with tools for math access?

The usability study~\cite{EuDML:UsabilityStudy:2010} conducted
interviews with mathematicians and essentially stated that
mathematicians didn't know how to use the offerings of mathematical
search interfaces. To get a better understanding we wanted to dig
deeper. In~\cite{Zhao:MIR:2008} {\name{Zhao}} concentrates on
user-centric and math-aware requirements for math search. The former
are based on mathematicians' specific information needs and search
behaviors, the latter are the needs for structured indizes by the
system. In contrast, we focus on eliciting attributions of existing
math search interfaces by mathematicians versus non-mathematicians. We
hope to learn what exactly sets mathematicians apart, since from this
knowledge we can deduce implications for future mathematical designs.

We decided on using repertory grid interviews as main
methodology to elicit evaluation schemes with respect to selected
math search interfaces (``{\bf{\mathUI}}'') and to understand how mathematicians
classify those {\mathUI}s. The main advantage of the method is its
semi-empirical nature. On the one hand, it allows to get deep insights
into the topic at hand through deconstruction and intense discussion of
each subject's idiosyncratic set of constructs and their resp. mapping
to the set of {\mathUI}s. On the other hand, the grids produced in
such \rgi sessions can be analyzed with a General Procrustes Analysis
to obtain statistically significant correlations between the elicited
constructs or the chosen {\mathUI}s. We used the \textsf{Idiogrid}~\cite{Grice:Idiogrid:2002} and the \textsf{OpenRepGrid}~\cite{OpenRepGrid} software for this.

Information search is not a single act, but a process with many strategies and options: {\myCitation{In fact, we move fluidly between models of ask, browse,
    filter, and search without noting the shift. We scan feeds, ask
    questions, browse answers, and search
    again.}}~\cite[p.7]{Morville:SearchPatterns:2010}. Therefore, we can consider the term ``search'' as an umbrella term for (at least) the following approaches:
\begin{description}
\item[Finding] = already knowing what one is looking for
  (\cite{NavarroEtAl:CognStrategiesSearching:1999,Shneiderman:ClarifyingSearch:1997}
  call it ``fact-finding'')
\item[Browsing] = getting an overview over a topic or an idea of a
  concept (\cite{NavarroEtAl:CognStrategiesSearching:1999} calls it
  ``exploration of availability'')
\item[Surfing] = surrendering to the links, drifting from one to
  another (see~\cite{WiseEtAl:SearchingVsSurfing:2009})
\item[Solving/Information Gathering] = creating a search plan, i.e., specifying a sequence
  of actions that achieves the solution of a problem
  (see~\cite[65ff.]{RussellNorvig:aiama95},~\cite{Kellar:InformationSeekingTasks:2007})
\item[Asking] = posing a question to find an answer
  (see~\cite{Taylor:ProcessOfAskingQuestions:1962})
\end{description}
Our question here is, what search approach is used with which
assessment attributes for what kind of math search tool? The answer
could enable us to design specifically for more math search approaches
by learning from the used ones.

We start out in Section~\ref{sec:study} with a description of the \rgi
study. In Section~\ref{sec:findings} we present the elicited interview
data and note the patterns that emerge from this data. The patterns
state interesting, prototypical attributions of mathematicians, which
separate the data gathered from the group of mathematicians from the
one of non-mathematicians. To demo the utility of such patterns, we
apply them in a discussion of an interesting, confusing evaluation of
two specific {\mathUI}s in Section~\ref{sec:example}. We conclude in
Section~\ref{sec:conclusion} by hinting at general design implications
for mathematical (search) interfaces based on the found set of
patterns.

\section{The Study}\label{sec:study}
The aim of our study was to find out what distinguishes mathematicians from
non-mathematicians when using a web interface for searching relevant content, here math
content. From the outset it was clear that observational methods wouldn't work as the
working context of a mathematician is typically neither restrained to certain locations
nor time slots. Surveys (or structured interviews) were out of question as the answers
require a deep insight of subjects into their own math search behavior, which
cannot be assumed in general.  Unstructured interviews could have been made use of to get
such deep insights, but we would either have to do
too many to be able to soundly interpret them or too few to draw general
conclusions. Finally,
the option of semi-structured interviews as methodology was discarded, since it became
clear in the first pilot study trials that mathematicians tend to describe ``truths'' and
``falsities''. In particular, they try to scrutinize the interview or interviewer and
manipulate the outcome towards what they think is the correct answer. Thus, the
interviewer has to trade her observational stance with a continuously sparring stance,
which hinders the process of gaining deep insights. 

In the end, we opted for the methodology of repertory grid interviews,
as they allow a semi-empirical analysis, and interviewees understand
quickly that they are not asked to decide on rights or wrongs.  The
\textbf{Repertory Grid Interview (\rgi)
  Technique}~\cite{HassenzahlWessler:RepGridTechnology00,Jankowicz:2003,Kelly:BriefIntroductionToPCT}
explores personal constructs, i.e., how persons perceive and
understand the world around them. It has been used as a usability/user
experience method to research users' personal constructs when
interacting with software artifacts
(see~\cite{HeideckerHassenzahl_RGTFuerAttraktivitaet_2007,HertzumEtAl:PersonalUsabilityConstructs2012,TanHunter:RGTinIT2002}
for examples). \rgi has the advantage that it can deliver valuable
insights into the perception of users even with relative low numbers
of study subjects (seeo~\cite{Kohlhase:HumanSpreadsheetInteraction:2013} for more details).

\begin{table}[ht]\centering\footnotesize\vspace{-2.em}
\begin{tabular}{|p{0.18\columnwidth}|p{0.58\columnwidth}|p{0.22\columnwidth}|}\hline
\tabhead{Element Name}&\tabhead{Short Description}&\tabhead{URL}\\\hline
\elemZBMathNew & {\small{ ``an {\textit{abstracting and reviewing service}} in pure and applied mathematics'' }} & {\scriptsize{\url{zbMath.org}}}\\\hline
\elemZBMathOld & {\small{ the former interface of \elemZBMathNew }} & {\scriptsize{not available}}\\\hline
\elemMathSciNet & {\small{ ``searchable {\textit{database of reviews, abstracts and bibliographic information}} for much of the mathematical sciences literature'' }} & {\scriptsize{\url{ams.org/mathscinet}}}\\\hline
\elemGoogleScholar & {\small{ ``search of {\textit{scholarly literature}} across many disciplines and sources'' }} & {\scriptsize{\url{scholar.google.com}}}\\\hline
\elemGoogle & {\small{ ``Search the {\textit{world's information}}, including webpages, images, videos and more'' }} & {\scriptsize{\url{google.com}}}\\\hline
\elemOffice & {\small{ the personal {\textit{office}} as math search interface}} & --- \\\hline
\elemTIB & {\small{ The {\textit{online catalogue}} of the Uni Hannover Library }} & {\scriptsize{\url{tib.uni-hannover.de}}}\\\hline
\elemVifamath & {\small{ ``The Virtual Library of Mathematics'' - a {\textit{meta online catalogue}}}} & {\scriptsize{\url{vifamath.de}}}\\\hline
\elemLibrary & {\small{ a {\textit{physical library}} known by the subject }} & --- \\\hline
\elemArxiv & {\small{ ``{\textit{Open e-print archive}} with over [\dots] 10000 [articles] in mathematics'' }} & {\scriptsize{\url{arxiv.org}}}\\\hline
\elemResearchGate & {\small{ ``a {\textit{network}} dedicated to science and research'' }} & {\scriptsize{\url{researchgate.net}}}\\\hline
\elemMathoverflow & {\small{ ``a {\textit{question and answer site}} for professional mathematicians'' }} & {\scriptsize{\url{mathoverflow.net}}}\\\hline
\elemColleagues & {\small{ personal {\textit{colleagues}} as math search interface }} & --- \\\hline
\elemMSCMap & {\small{ ``accessing math via {\textit{interactive maps}}'' based on an MSC metric }} & {\scriptsize{\url{map.mathweb.org}}}\\\hline
\elemCatchup & {\small{ an interface for {\textit{catching up}} with the newest articles in math }} & {\scriptsize{\url{arxiv.org/catchup}}}\\\hline
\elemFormulaSearch & {\small{ ``allows to search for {\textit{mathematical formulae}} in documents indexed in zbMath'' }} & {\scriptsize{\url{zbmath.org/formulae}}}\\\hline
\elemBibliography & {\small{ a {\textit{bibliography} as math search interface}}} & \\\hline
\end{tabular}\vspace{0.5em}\\
\caption{The \rgi Elements in the Study}\label{tab:elements}\vspace{-3.0em}
\end{table}
\subsection{The \rgi Elements}
As we want to cover a broad range of different types of math search
interfaces we opted for a set of 17 {\mathUI}s as {\bf{\rgi elements}}
-- ranging from standard {\mathUI}s like ``Zentralblatt Mathematik
(zbMath)'' or ``MathSciNet'' via social media platforms like
``mathoverflow'' to scientific prototypes like the ``MSC map''
interface (MSC = Math Subject Classification, see~\cite{MSC-SKOS}). To
avoid being limited to digital {\mathUI}s, we included traditional
search situations like asking colleagues or personal office spaces as
well. Table~\ref{tab:elements} summarizes the 17 elements used in the
{\rgi}s and gives short descriptions -- the ones from their websites
where available -- and their web addresses if applicable.  Note that
wikis (e.g., ``Wikipedia'' or ``PlanetMath'') were excluded as the
tension between searching for articles versus encyclopedia entries was
perceived problematic in the pilot study, so we opted for the
former. As we were only interested in the search behavior of
mathematicians we disregarded mathematical software whose main task is
computation or verification.

\subsection{The \rgi Set-Up}
At the beginning of each interview the interviewer introduced the
interviewee to all {\mathUI}s based on print-outs. Both the home page
with its search facilities and the search result pages were
discussed. The front page of each print-out presented the homepage
initialized with the phrase ``Cauchy sequence'' in the search box if
applicable.  The back page displayed the search result wrt to this
query. For {\mathUI}s with special features extra pages were
attached. For {\elemFormulaSearch} the {\LaTeX} query corresponding to
$?a_{?n}\in\mathtt{N}$ was used.

An \rgi interview iterates the following process until the
interviewee's individual construct space seems to be exhausted:
\begin{compactenum}[\em i.]
\item The interviewee randomly chooses three \rgi elements.
\item He declares which two of the three elements seem more similar.
\item He determines the aspect under which these two are more similar
  and the aspect under which the third one is different. Those aspects
  are the ``{\bf{poles}}'' of an interviewee-dependent evaluation
  dimension, the so-called ``{\bf{construct}}''.
\end{compactenum}
To get a sense of what the users consider important properties of
{\mathUI}s, we extended this set-up by encouraging most interviewees
to judge the ``fitness'' of each {\mathUI} for mathematical search.
As is typical with {\rgi}s, the interviews were very
intense. Therefore, the findings are not only based on the
actual data elicited in the {\rgi} but also on the deep
discussions taking place during each interview.

\subsection{The {\rgi} Data}
We conducted interviews with 24 people, all of which were interested
in accessing math on the web. Out of these, 18 had a degree in
mathematics. For the final analysis we decided to use 22 {\rgi}s:
interviews with a group of 11 professional mathematicians working in a
scientific environment (``\Math''), a group of 5 content experts for
mathematical information (``\zbMath''), and a group of 6
non-mathematicians (``\noMath''). Only 3 of the
participants were female.

Each interview took between 1.75 and 3 hours, in which an average of 4
constructs were elicited. The \Math group created 50 constructs,
{\zbMath} reported 28 constructs and \noMath 29 constructs.  The
rating scale for these 107 elicited constructs was a 7-point Likert
scale.

\section{Findings}\label{sec:findings}
As already mentioned, the {\rgi} method is semi-empirical. This means
that there will be a quantitative and a qualitative analysis of the
data gathered. Due to space limitations we will focus on presenting and interpreting the most interesting, statistically significant quantitative results in form of dendrograms and qualitative results in form of patterns. Note that here, the theory emerges from the data, thus, it provides us
with patterns but not with proofs.

With the {\bf{Generalized Procrustes Analysis (GPA)}} method
(see~\cite{Gower:GeneralizedProcrustesAnalysis_1975})
3-dim\-ensional data matrices can be analyzed with a multivariate
statistical technique. In particular, in our RGI we can compare the
individual (dim 1) natural lan\-guage con\-structs (dim 2) rated on
our fixed set of {\mathUI}s (dim 3). We conducted a GPA
with {\textsf{Idiogrid}} for each data set and refer
to~\cite{Koh:FramingsOfInformation:2013} for a de\-tailed description
of an analoguous GPA procedure. To provide a shared set of (virtual)
standard constructs on which the individual ratings of the \rgi
elements of each interviewee can be compared, the GPA method produces ``{\bf{abstract constructs}}'' of the form ``Con\_i -
ConOpo\_i'' with poles ``Con\_i'' and ``ConOpo\_i''. 

\begin{figure}[h]\vspace*{-2em}
  \includegraphics[width=\columnwidth]{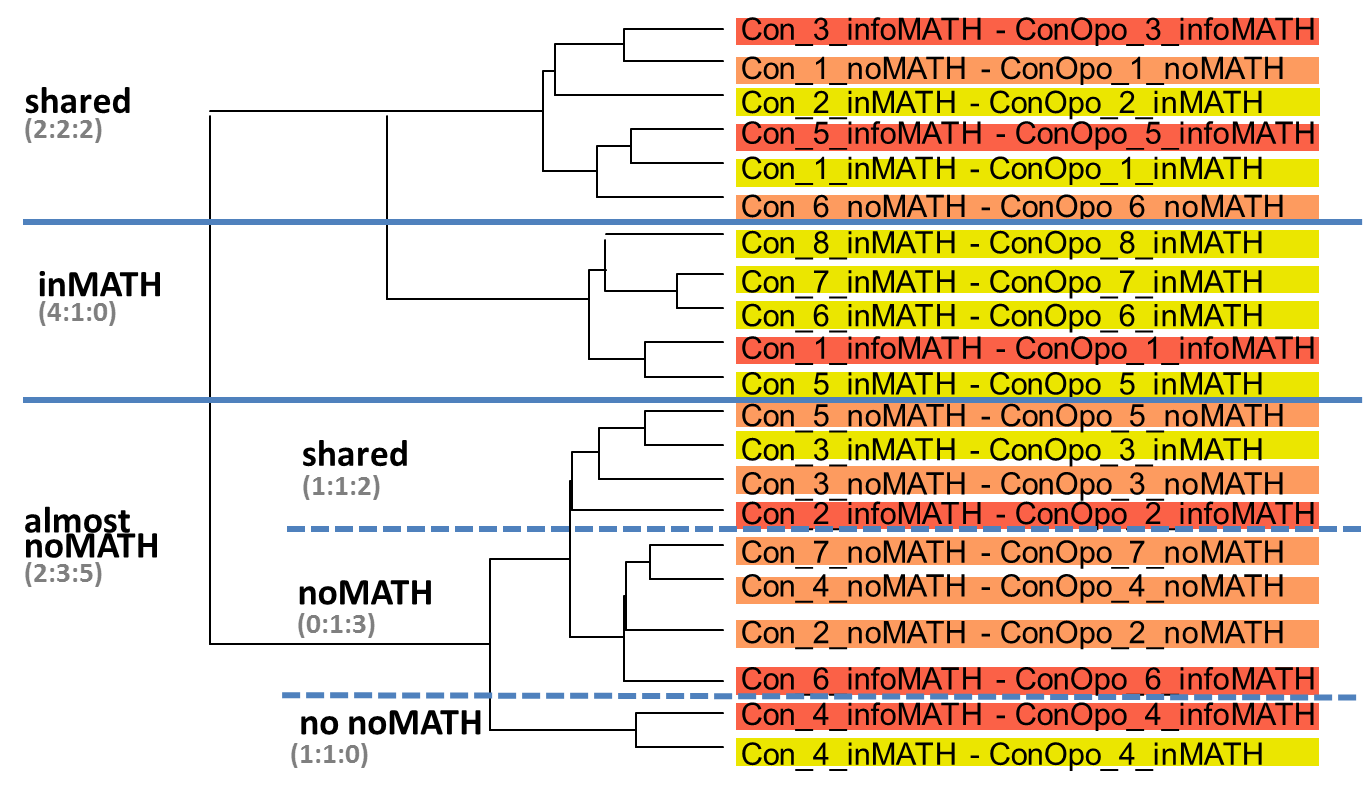}\vspace*{-1em}
  \caption{Dendogram of the Abstract Construct Clusters
    (wrt. Euclidean distance and Ward clustering) of \Math, \zbMath
    and \noMath: we can clearly discern a ``common'' cluster, which is
    equally shared by all three, a strong \Math cluster and a fairly
    strong \noMath cluster.}\vspace{-1.0em}
  \label{fig:distinction}\vspace{-1.0em}
\end{figure}
Based on a pre-study we suspected a distinction of the interviewee
group not only into mathematicians and non-mathematicians, but into
research mathematicians, mathematics practitioners and
non-mathematicians. Therefore we compared the
element evaluations of the \Math, \zbMath, and \noMath group. We subjected
the union of the group-specific sets of abstract constructs to a
cluster analysis run by \textsf{OpenRepGrid} resulting in the
dendrogram in Fig.~\ref{fig:distinction}. Recall that
\textbf{dendrograms} are a visual representation of correlation
data. Two constructs in Fig.~\ref{fig:distinction} are closely
correlated, if their scores on the \rgi elements are similar. The
distance to the next upper level of two constructs/groups of
constructs {\emph{indicates}} this relative closeness. Please note
that we left out the scale in the dendrograms, as we are not
interested in the absolute numbers, only in their relative
groupings. This also means, that we won't use arguments in our
discussion of findings based on this scale. Nevertheless, we can for
example, conclude from Fig.~\ref{fig:distinction} that \con{6}{\Math}
and \con{7}{\Math} are the most correlated constructs. For the
conversion of \textsf{Idiogrid} data to \textsf{OpenRepGrid} data we
developed the according software.

The interview data seen in Fig.~\ref{fig:distinction} indeed suggest a difference
between how people in the \Math, \zbMath and the \noMath group think about {\mathUI}s.
The \zbMath interviewees' point of view lies between the one of \Math and \noMath
subjects. In particular, there are \zbMath abstract constructs in every cluster and there
is no cluster dominated by the \zbMath abstract constructs. As this user group dilutes
possible similarities or dissimilarities wrt the user group in focus -- the professional
mathematicians -- we further on only analyzed the \Math and \noMath data in depth. From here
on we will call \Math members ``mathematicians'' and \noMath members
``non-mathematicians''.

\begin{figure}\vspace*{-2em}
\includegraphics[width=\columnwidth]{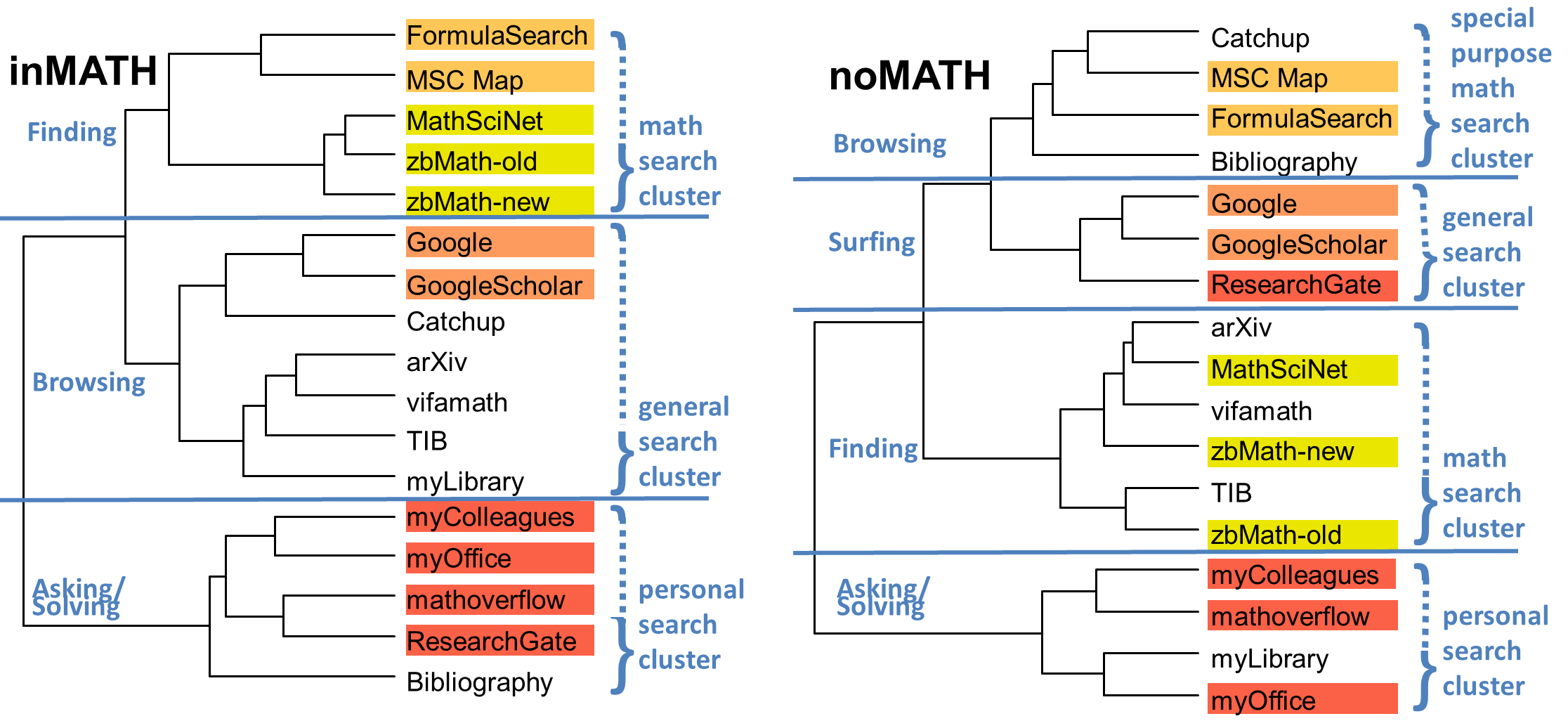}\vspace*{-1em}
\caption{Cluster Dendrograms of {\mathUI} Elements for \Math and
  \noMath}\label{fig:elementCluster}\vspace*{-1em}
\end{figure}
Fig.~\ref{fig:elementCluster} gives a visualization of the element clusters of group
\Math resp. \noMath as dendrograms. The difference between the clusters is evident; we
will elaborate the interpretations in the next paragraphs.

There are three main element clusters for \Math in
Fig.~\ref{fig:elementCluster}. Clearly, one of these contains the {\mathUI} elements
whose main purpose it is to find mathematical content (``{\bf{math search cluster}}'').
In the math search cluster both, {\elemFormulaSearch} and {\elemMSCMap}, are innovative
mathematical services, nevertheless they are identified as being most similar to the
standard {\mathUI}s {\elemZBMathNew}, {\elemMathSciNet} and {\elemZBMathOld}. This shows
that \myPatterns[pat:Familiarity]{Mathematicians do not assess {\mathUI}s based on
  familiarity.}

Another cluster includes all {\mathUI} elements that provide a
personal touch in the search process (``{\bf{personal search
    cluster}}''). Here, the term ``personal'' labels the
interactive adaptation and customization of the search or search
results in a process driven by human interactions. In the interviews it became
quite clear, that anything involving human beings or communities was
highly distinctive and predominantly highly appreciated. So
\myPatterns[pat:Community]{Mathematicians trust human and community resources.}

Note that we don't mean a naive trust here, but a trust given the
sensible precautions. Even though the element clusters of the \noMath
interviewees also include a personal search cluster (see
Fig.~\ref{fig:elementCluster}), the elements {\elemBibliography} and
{\elemResearchGate} are missing and replaced by {\elemLibrary}. The
\Math participants explicitly commented that they don't have
confidence in the librarians' expertise in math. Interestingly,
mathematicians showed a lot of skepticism wrt {\elemResearchGate} but
not because they could not rely on the links the {\elemResearchGate}
members would provide, but rather because they mistrusted
{\elemResearchGate}'s competence in judging the relevance of links.
An indication of this is also given by the well-known observation that
mathematicians like anecdotes about fellow mathematicians like no
other community of practice.

The third cluster groups the remaining elements. Noticeably
{\elemGoogle} and {\elemGoogleScholar}, which mathematicians nowadays
use heavily for mathematical searches, are in this
cluster. Nevertheless, these elements
are not specific to math search, therefore we label this cluster as
the ``{\bf{general search cluster}}''.

According to {\name{Zhao}}'s usability study in~\cite{Zhao:MIR:2008},
mathematicians use {\myCitation{three main approaches: general keyword
    search, browsing math-specific resources and personal contact.}}
This can also be seen in our three clusters for the \Math
group.

For the \noMath element clusters we only want to point out that the
clusters are indeed very different from the ones in the \Math
dendrogram. For example, for mathematicians the {\mathUI}s
\elemMathSciNet, \elemZBMathNew and \elemZBMathOld correlate the
highest, whereas for non-mathematicians each of them correlates more
with a different element than with each other. The only similarities
seem to be the obvious correlation between \elemGoogle and
\elemGoogleScholar, and the same very high correlation distance
between the personal search cluster and the others.


\begin{figure}[h]
\includegraphics[width=\columnwidth]{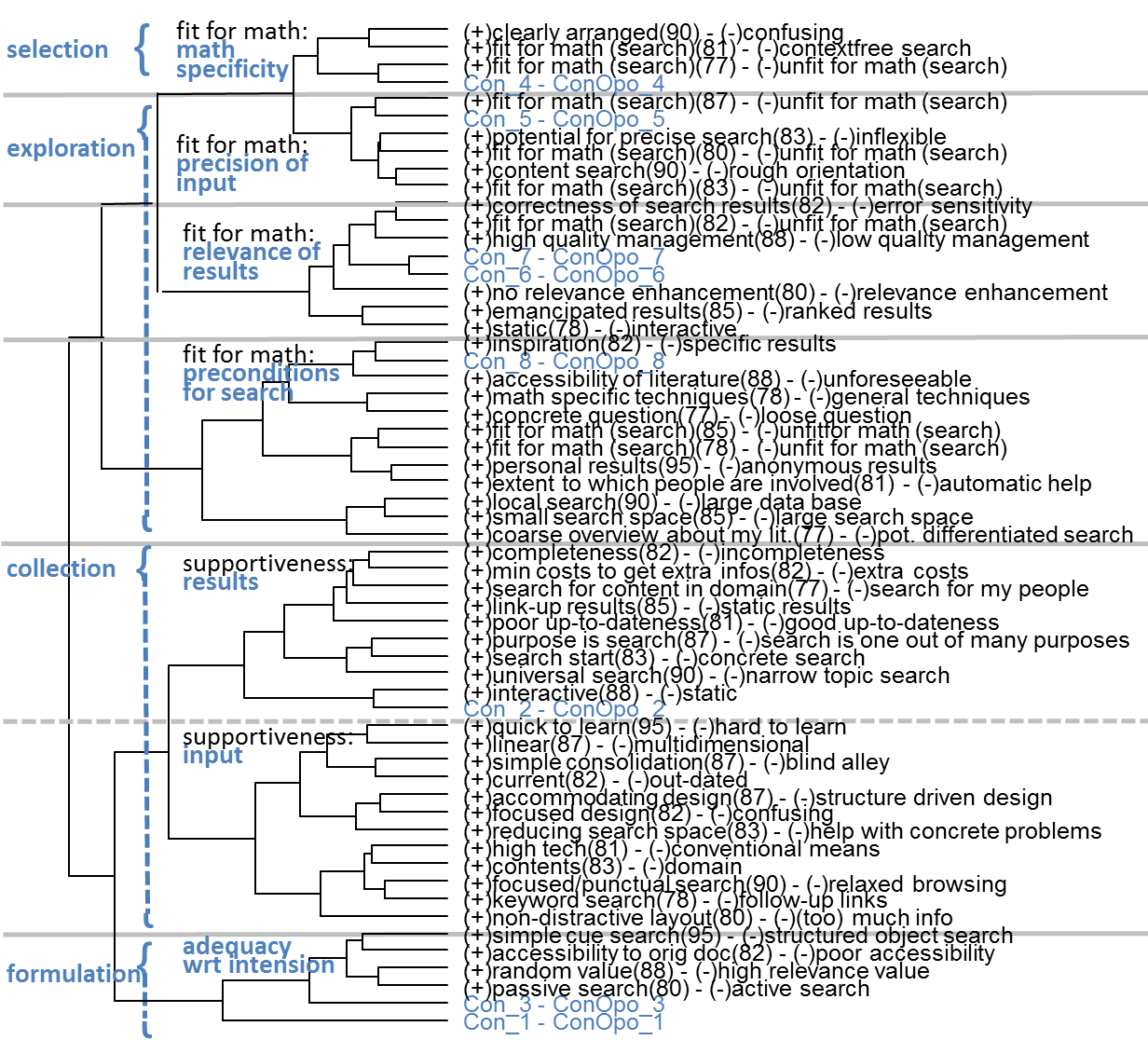}\vspace*{-1em}
 \caption{Cluster Dendrogram of Construct Clusters in \Math. The first
   two levels of the dendrogram were contracted for a more readable
   image. Moreover, the numbers in parentheses in each construct
   encode the individual interviewee issuing
   it.}\label{fig:constructClustersMath}\vspace*{-2em}
\end{figure}
For a more precise qualitative analysis consider the dendrogram in
Fig.~\ref{fig:constructClustersMath}. First we decided on fitting
catego\-ries/sub\-ca\-te\-go\-ries for each cluster. We looked, for instance, at the first
main cluster and decided on the category ``fit for math''. Then we elaborated on its four
subclusters, e.g., for the fourth cluster we selected ``preconditions for search'' as a
subcategory. Note that there are blue-colored abstract constructs ``Con\_i -
ConOpo\_i'' among the constructs. We can interpret them now as characteristic constructs
of the corresponding major subcluster, so we associate each abstract construct with its
subcategory. Out of convenience, we call them by their explicit pole name together with
the corresponding data set, thus we say for example, ``\con{4}{\Math} `means' math
specificity''.

According to \name{Kuhltau et al.}
in~\cite{KuhltauEtAl:InformationSeekingRevisited:2008,Kuhltau:SeekingMeaning:2004} the
information search process can be described by a six-phase framework consisting of
{\emph{initiation}} (prompting a search), {\emph{selection}} (identifying information
needs), {\emph{exploration}} (pondering available tools and thus search strategies),
{\emph{formulation}} (formalizing search queries), {\emph{collection}} (gathering
information and goal-oriented cherry picking in search results), and {\emph{search
    closure}} (giving up on the search). In our study we are not interested in the entire
search process, but in the interactions with the user interface, so we focus on the
iterative acts of selection, exploration, formulation and collection. In these phases a
user seeking information translates a search intension into a query or series of queries
optimizing for the relevance of the final collection of search results. Interestingly, the
four phases are mirrored in the construct clusters of the \Math group (see
Fig.~\ref{fig:constructClustersMath} on the left).

\begin{figure}\center\vspace{-2.5em}
\includegraphics[width=0.8\columnwidth]{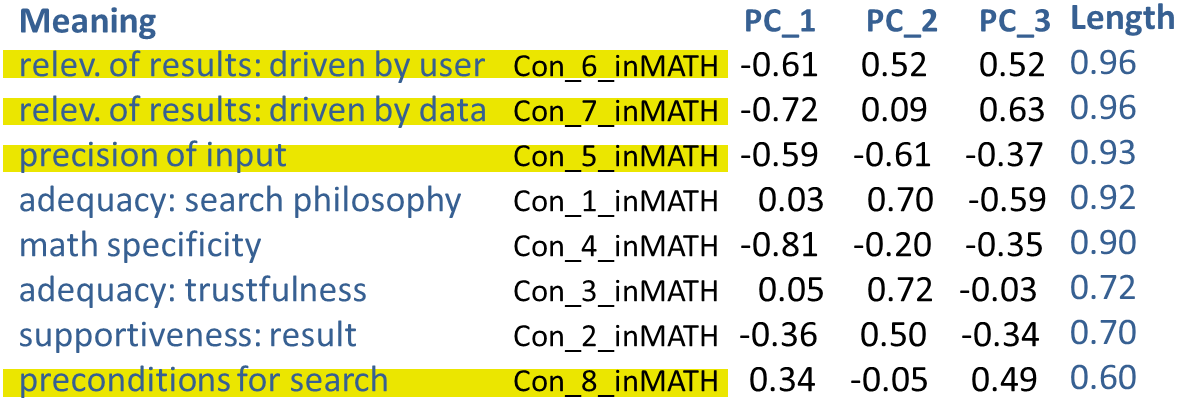}\vspace*{-1em}
 \caption{Abstract Construct Ranking in \Math via Structure Coefficients}\label{fig:structureCoefficientsMath}\vspace*{-2.0em}
\end{figure}
\begin{figure}\center\vspace{-2.0em}
\includegraphics[width=0.8\columnwidth]{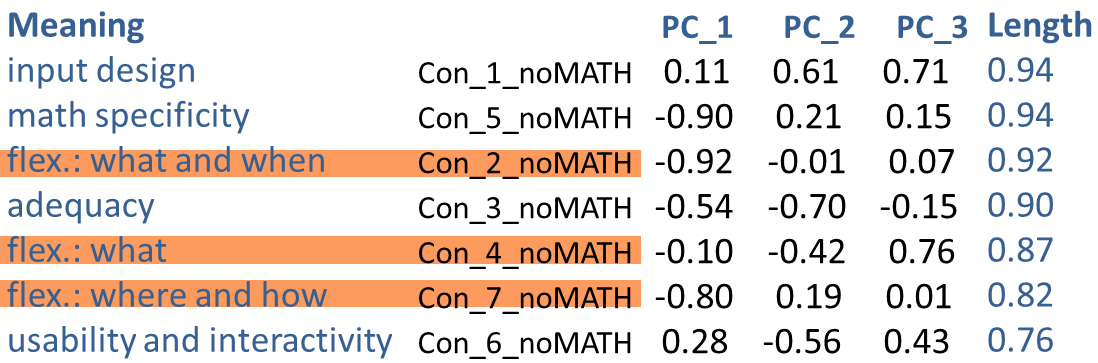}\vspace*{-1em}
 \caption{Abstract Construct Ranking in \noMath via Structure
   Coefficients}\label{fig:structureCoefficientsNoMath}\vspace*{-2.0em}
\end{figure}

To obtain a {\bf{ranking for the abstract constructs}} consider the
structure coefficients of the abstract constructs (wrt their ratings
on the three main principal components $PC_i$) for each interviewee
group in Fig.~\ref{fig:structureCoefficientsMath}
and~\ref{fig:structureCoefficientsNoMath}. The Euclidean {\bf{length}} of the
resp. 3-dimensional construct vector indicates its construct's
relevance. To distinguish between two abstract constructs that are in
the same subcluster in Fig.~\ref{fig:constructClustersMath}, we
compare their structure coefficients. For any distinctive deviation we
take a closer look in the biplots for the resp. PC-dimension and
elaborate on its meaning.

It is obvious that the rankings of the \Math group are distinct from
the ones of the \noMath group. What mathematicians care about the most
is the relevance of the search result with respect to their search
intension. So they seek interfaces and databases that allow them to
formulate precisely that in accordance with the respective search
philosophy they want to apply ({\con{6}{\Math}}, {\con{7}{\Math}},
{\con{5}{\Math}}, {\con{1}{\Math}}). As this describes a search
process that enables the user to find exactly what he is looking for,
we have \myPatterns[pat:Finding]{Finding is the primary mathematical
  search task.}
 
Note that the math search cluster of the \Math group in Fig.~\ref{fig:elementCluster}
also has a clear focus on ``finding''. For the \noMath math search cluster this is much
less clear, e.g., the {\elemVifamath} \mathUI, which concentrates on collecting mathematical
information (from legacy math articles to images of mathematicians), but not on its
findability, thus mimicking a physical math library without noticeable presence of other
people. The interviewer observed that interviewees aligned the distinct kinds of search
like finding, browsing or solving with the clusters, but that the evaluation of search
activities was different for mathematicians and non-mathematicians. The former had a clear
preference for finding, followed by browsing and solving/asking, and even a hint of
rejection for surfing. In contrast, the \noMath participants indicated a preference for
browsing and surfing, followed by solving/asking and finally finding. Note that the
position of ``finding'' in this ranking may be well due to the fact, that only one
participant in the \noMath group worked in a scientific environment.

It is conspicuous that even though there was an obvious \mathUI
cluster with respect to ``people'' (the personal search cluster) for
the \Math group in Fig.~\ref{fig:constructClustersMath}, there is no
appreciation of ``socialness'' in their ranked list of constructs in
Fig.~\ref{fig:structureCoefficientsMath}. In particular,
mathematicians distinguish certain {\mathUI}s, i.e., the tools, as socially driven, but as
professionals they do not appreciate ``socialness'' as a value per
se in their evaluation schemes. In the theory of ``Communities of Practice
({\bf{CoP}})''~\cite{SituatedLearning}, {\bf{practices}} are not only typical
customs shared within a community, but they are tools that define the
community.  Whereas in other CoPs
social interaction is a tool for achieving {\emph{social}} bindings,
in the mathematical CoP, social interaction is a tool for doing
mathematics, i.e., it is a mathematical practice.  Therefore, we note
that \myPatterns[pat:SocialInteractionTool]{Mathematicians appreciate
  social interaction as a {\emph{mathematical}} tool. In particular,
  it is a mathematical practice to collaborate and exchange feedback.}
In this sense, we confirm {\name{Brown}}'s dictum
in~\cite{Brown:InformationSeekingScientists:1999} that mathematicians
may rely more heavily on their social network than other disciplines.

Let us recall from Fig.~\ref{fig:distinction}, that some {\mathUI}
elements were in a subcluster shared by all three user groups. That
is, with respect to these constructs the \Math, \zbMath as well as
\noMath interviewees agreed on the evaluation of the given
        {\mathUI}s. In particular, {\mathUI} scores correlate on
        \con{3}{\zbMath} (``usability''), 
       {\con{1}{\noMath}} (``input design''), 
        \con{5}{\zbMath} (``simple design''),
{\con{6}{\noMath}} (``usability and interactivity''), and {\con{2}{\Math}} (``supportiveness:result''), {\con{1}{\Math}} (``adequacy: search philosophy''). 

We note the different flavor of the non-\Math constructs versus the \Math
constructs. Where the former aim for design aspects, the latter are only concerned with
fitness of the {\mathUI} for achieving the search intension. It becomes even clearer if we
consider the phrasing ``usability'' in the non-\Math group and ``supportiveness'' in the
\Math group: Whereas usability is a neutral measure for all kinds of qualities while using
an object, supportiveness is a task-oriented requirement in the use-flow of a human
person.  The media-theoretic difference is that the first doesn't tell us anything about
whether the user adopts a {\mathUI} as a mere tool or as a medium (in the sense of
{\name{McLuhan} as {\myCitation{any extension of the human body [\ldots] as a side-effect
      of a technology}}~\cite[564]{McLuhan:ExtensionsOfMan:2003}}, i.e., a technology that
empowers its users): \myPatterns[pat:Medium]{Mathematicians aim at adopting a search tool
  as a medium.}  One consequence is that once they have adopted it as a medium, they won't
easily change to other media.  Not surprisingly, this shared construct cluster also
supports a long-standing belief that \myPatterns[pat:Function]{Mathematicians appreciate
  function over form.}

Even though the {\mathUI} elements' scores were highly correlated in
the shared cluster, their respective conceptualization can still
disagree.  To understand the conceptualization, we look at the
meanings of the distinct constructs and the location of an element wrt
constructs. For instance, for mathematicians {\elemGoogleScholar}
enables a top-down approach (as search philosophy =\con{1}{\Math}) by
using a very general technique of ranking the search results (high
supportiveness for presenting search results =\con{2}{\Math}), but
offers a very textual input design (=\con{1}{\noMath}) and a
medium-rated effectiveness (as part of usability = \con{6}{\noMath})
for non-mathematicians. Here, note that the evaluation by the
mathematicians concerns the outcome, whereas the non-mathematicians
rather assess it by the input. This argument can be made more
generally, as the resp. construct clusters for the \Math
resp. \noMath groups favor the result resp. the input. Interestingly
one cluster category of the \Math group didn't make it into the
consensus grid. In particular, the category ``supportiveness of
input'' has no representative among the abstract constructs of
\Math. We conclude \myPatterns[pat:Outcome]{Mathematicians care
  more for the outcome than the input.}  This also means that
mathematicians seem to be willing to trade input hardships (like more
complex interfaces) for output satisfaction (i.e., having perfect
{\bf{precision}} -- all found results fit the search intension-- and
{\bf{recall}} -- all fitting results were found).
 
In Fig.~\ref{fig:distinction} we observe that the constructs
{\con{5}{\Math}}, {\con{6}{\Math}}, {\con{7}{\Math}}, and
{\con{8}{\Math}} are part of an abstract construct cluster only
containing \Math constructs; they are the enhanced (yellow colored)
constructs in Fig.~\ref{fig:structureCoefficientsMath}.  Here, the
math interfaces scored similarly according to the attributes
``relevance of results: driven by user'', ``relevance of results:
driven by data'', ``precision of input'', and ``preconditions of
search''. Thus, we can interpret that a mathematical search interface
that empowers the user by enabling him to fine-tune the search query
is considered to strongly improve the relevance of the result.  This
interpretation is supported by Pattern~\myPattern{pat:Outcome}, thus
we note that \myPatterns[pat:Empowerment]{Mathematicians want to be
  empowered in the search process.}

Moreover, mathematicians obviously realize that this precision comes at a cost: the
underlying data have to be structured enough. Therefore, if the data do not allow such a
fine-tuning right away, they are willing to iteratively refine their query themselves.  A
direct consequence seems to be that mathematicians want as much support in formulating a
search query as they can get.
Whereas non-mathematicians will agree that Pattern~\myPattern{pat:Outcome} is different
from their own approach, wrt the above consequence their attitude might be different as
the pattern describes the disregard of input facilities by mathematicians and the latter
the total investment of time and energy towards satisfying the search
intension. We already
observed that this is a cluster of elements marked by mathematicians only. That is, this
kind of evaluation scheme didn't occur to non-mathematicians, thus it isn't a dominant
one.

For Pattern~\myPattern{pat:Finding} we argued with the abstract construct ranking within
Fig.~\ref{fig:structureCoefficientsMath}. Interestingly, three of the four first ranked
items in that list belong to the uniquely mathematical cluster. The fourth one
(\con{8}{\Math}, ``preconditions for search'') occurred in the \Math group only, that is,
it discriminates between mathematicians' and non-mathematicians' search behavior ever
more. As mathematicians take the preconditions for search into account in the exploration
phase of an information search process, they value their anticipation of the search
outcome. This has two consequences: \myPatterns[pat:Transparency]{Mathematicians base
  their information search process on transparency of the search result.}

Additionally, if they put a lot of thought into the exploration phase, they expect to be
rewarded by a good search result. So we hold \myPatterns[pat:Expectations]{Mathematicians
  expect to find meaningful information in the search result.}  In the interviews, it was
striking how much awe {\elemGoogle} evoked. Pattern~\myPattern{pat:Expectations} solves
this riddle: Considering the low amount of work to be invested in the exploration phase,
the expectations towards the search results are really low. Therefore, the relevance of
{\elemGoogle} searches amazes mathematicians tremendously.
 


\section{Understanding the Mathematical Perspective on {\mathUI}s: an
  Example}\label{sec:example}
To see the utility of the elicited patterns, we will now discuss the {\mathUI}s \elemMathSciNet and \elemZBMathNew under a mathematical perspective, which is informed by our elicited patterns. 

\begin{wrapfigure}{r}{7.2cm}\vspace*{-2em} 
\includegraphics[width=0.6\columnwidth]{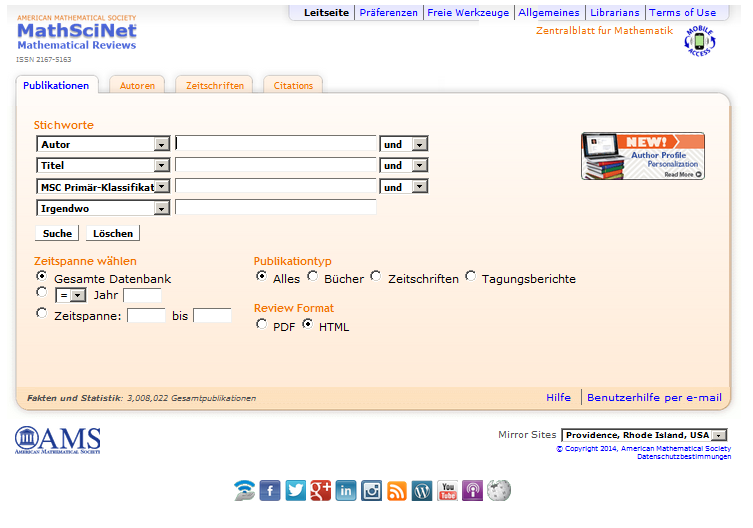}\vspace*{-1em}
 \caption{\mathUI of {\elemMathSciNet}}\label{fig:MSN}\vspace*{-2.5em}
\end{wrapfigure}
\begin{wrapfigure}{l}{7.2cm}\vspace*{-2.0em} 
\includegraphics[width=0.6\columnwidth]{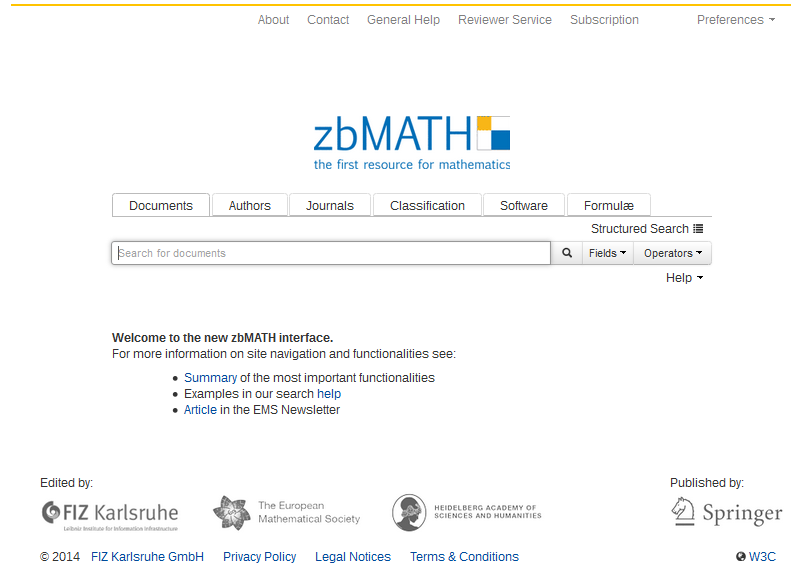}\vspace*{-1em}
 \caption{\mathUI of {\elemZBMathNew}}\label{fig:zbMath}\vspace*{-2.0em}
\end{wrapfigure}

Let us start with {\elemMathSciNet} as seen in Fig.~\ref{fig:MSN} and {\elemZBMathNew}
shown in Fig.~\ref{fig:zbMath}.  From above we know that mathematicians don't discern
between {\elemMathSciNet} and {\elemZBMathNew}. This immediately rai\-ses the question why
this might be the case. Evidently both layouts use a lot of vacuity to focus the users' attention
and use bright colors sparingly. But we know because of Pattern~\myPattern{pat:Function},
that the form is not important to mathematicians, so the reason for their alignment cannot
stem from these observations. 
Unfortunately, at first glance the similarity of the start
page already ends here: {\elemZBMathNew} provides a simple search, i.e., a one-step
search, {\elemMathSciNet} a multi-dimensional, structured search.
Moreover, {\elemZBMathNew} offers innovative extra services
like mathematical software search and formula search,
{\elemMathSciNet} an extra citation service. The former offers inline
search fields to specify the search. The latter provides social media
connections. If we look at the search result page of each, we will find that there are as many differences.

Now let us take a closer look, for example, at the difference between
{\elemZBMathNew}'s simple search and {\elemMathSciNet}'s structured
search. We know because of Pattern~\myPattern{pat:Outcome}, that
mathematicians value the outcome higher than the input. Therefore, as
{\elemZBMathNew} offers not only the functionality of
{\elemMathSciNet}s multi-dimensional search via inline search fields
in the simple search but also choicewise a link to a structured
search, the functionality seems to be the same for mathematicians. The
input inefficiencies can be neglected, the potential outcome is the
same. 

 What about the clear differences in functionality in these
 {\mathUI}s? Note that the social media links weren't recognized once
 in the interviews with the mathematicians, which also fits
 Pattern~\myPattern{pat:SocialInteractionTool}, stating that they
 appreciate the communities' practices, but not the links to a
 community themselves.  Then, it seems rather evident that
 {\elemZBMathNew} offers {\emph{more}} functionality, as
 {\elemMathSciNet}'s extra functionality consists only of the citations
 index. So shouldn't Pattern~\myPattern{pat:Empowerment} kick in and
 lead to a distinctive perception of both systems? 

We can counter-argue with two patterns. On the one hand,
Pattern~\myPattern{pat:Finding} tells us that finding is the major
kind of search a mathematician is conducting. The additional services
{\elemZBMathNew} provides the user with are essentially no services
that support finding, they rather support browsing. This is clear for
the mathematical software search. The facetted search with its
abilities to refine a search in the process also supports browsing
behavior explicitly. In contrast, the formula search feature was
designed for finding, but in the interviews, mathematicians indicated
that they simply don't believe in the finding capability of the
software (unfair as it is). In~\cite{Zhao:MIR:2008}, interestingly, a
similar phenomenon was observed. The underlying reason for this
disbelief could lie in Pattern~\myPattern{pat:Medium}, namely that
they have adopted {\elemZBMathNew} as a medium, and that uses string
search. Therefore, their conceptualization of this service doesn't fit
yet and is a challenge to change.

On the other hand, {\elemZBMathNew} is rather new. The older version didn't have as many
relevant extra features as this new one. Thus, Pattern~\myPattern{pat:Medium} strikes
again. Quite a few interviewees reported that they use {\elemMathSciNet} and even when
they became aware that {\elemZBMathNew} has more to offer now, they didn't mention any
intention to change over.\medskip

We can summarize that the patterns help us understand the perception of mathematicians
much better. This new-found understanding in turn triggers new design challenges and
ultimately better, more math-oriented designs.

\section{Conclusion}\label{sec:conclusion}
We have presented an \rgi study that was concerned with mathematical
search interfaces. To be able to understand the idiosyncracies of
mathematicians, we interviewed mathematicians as well as
non-mathematicians, with a focus on the former. From the quantitative
data and its qualitative interpretation several patterns emerged:
\begin{description}
\item [P~\myPattern{pat:Familiarity}] {\myCitation{Mathematicians
      do not assess {\mathUI}s based on familiarity.}}
\item [P~\myPattern{pat:Community}] {\myCitation{Mathematicians trust human and community resources.}}
\item [P~\myPattern{pat:Finding}] {\myCitation{Finding is the
      primary mathematical search task.}}
\item [P~\myPattern{pat:SocialInteractionTool}]
  {\myCitation{Mathematicians appreciate social interaction as a
        mathematical tool. In particular, it is a mathematical
        practice to collaborate and exchange feedback.}}
\item [P~\myPattern{pat:Medium}] {\myCitation{Mathematicians aim
      at adopting a search tool as a medium.}}
\item [P~\myPattern{pat:Function}] {\myCitation{Mathematicians
      appreciate function over form.}}
\item [P~\myPattern{pat:Outcome}] {\myCitation{Mathematicians
        care more for the outcome than the input.}}
\item [P~\myPattern{pat:Empowerment}] {\myCitation{Mathematicians
      want to be empowered in the search process.}}
\item [P~\myPattern{pat:Transparency}]
  {\myCitation{Mathematicians base their information search process
        on transparency of the search result.}}
\item [P~\myPattern{pat:Expectations}]{\myCitation{Mathematicians
  expect to find meaningful information in the search result.}}
\end{description}
With these patterns many design issues for {\mathUI}s can be understood and elaborated on much deeper now.

For instance, {\name{Libbrecht}} posed
in\cite{Libbrecht:TooPreciseTopicQueries:2013} the question whether
(mathematical) search queries may become too precise (so that the
search result becomes too small). But this question does only make
sense for browsing queries not for finding queries. The
Pattern~\myPattern{pat:Empowerment} suggests that the solutions should
be finetuned to the distinct kind of searches. If that is not
possible, the default case should be ``finding'' because of
Pattern~\myPattern{pat:Finding}.

Pattern~\myPattern{pat:Medium} indicates that a change from one tool
to another is not easily done by mathematicians. In particular, a
change of media will only occur if the innovation is disruptive, a
mere incremental innovation won't suffice. Therefore, phrasing a major
change (like the one from {\elemZBMathOld} to {\elemZBMathNew}) as a mere
update won't convince mathematicians to switch, and because of
Pattern~\myPattern{pat:Outcome}, neither will an announcement of
change that essentially points to the new Google-like layout of the
homepage.

Moreover, our data suggest that the search approach ``finding'' is
used by mathematicians predominantly when interacting with elements
from the math search cluster, ``browsing'' when interacting with
{\mathUI}s in the general search cluster and ``solving/asking'' when
using elements in the personal search cluster. Thus, we can look for
the properties of the resp. cluster to extend {\mathUI}s by more
search approaches. Note that \elemGoogle is best-known for its
browsing qualities, only for specific kinds of queries it is now also
successful in finding. Under this aspect \elemGoogle is also often
used by mathematicians.

Our future work is concerned with general design implications based on
the foundational work conducted in this paper. For example, one simple
consequence concerns the development process of math user interface
development: Specify the user group of your math service beforehand
and appreciate the credo of ``participatory design'' that strongly
admonishes developers to acknowledge the fact that {\emph{``You are
    not the user!''}}. In our study, e.g., the mathematics
practitioners turned out to be different from the professional
mathematicians.  Another consequence might be that we should make
mathematicians more effective by supporting their
{\emph{interventions}} in formulating a search query.

Finally, we like to note that the \rgi methodology -- even though strenuous at times -- seems to be a worthy methodology for use
  with mathematicians.

\subsubsection*{Acknowledgement}
I thank all my interviewees for their motivation and patience with the
\rgi method. Moreover, I appreciated the supportive work environment
at zbMath, especially discussions with Wolfram Sperber.  This work has
been funded by the Leibniz association under grant SAW-2012-FIZ. The final publication is available at {\url{http://link.springer.com}}.




\renewcommand\baselinestretch{.97} \printbibliography
\end{document}